\documentclass[letterpaper, 10 pt, conference]{cssconf}

\IEEEoverridecommandlockouts                              % This command is only needed if 
\usepackage{url}
\usepackage{tikz}
\usepackage{float}
\usepackage{multirow}

\title{\LARGE \bf
Characterization of a Multi-User Indoor Positioning System\\ 
Based on Low Cost Depth Vision (Kinect) for Monitoring\\ 
Human Activity in a Smart Home}

\author{Lo\"ic~Sevrin,
        Norbert~Noury, %~\IEEEmembership{Senior Member,~IEEE,}
        Nacer~Abouchi, %~\IEEEmembership{Member,~IEEE,}
        Fabrice~Jumel,
        Bertrand~Massot,
        and~Jacques~Saraydaryan% <-this % stops a space
\thanks{ %This paper was submitted on February, 16th.
This work was supported by a special grant from the Institute of Nanotechnologies of Lyon (INL).}% <-this % stops a space
\thanks{L. Sevrin is with the University of Lyon, INL, UCBL 
{\tt\small loic.sevrin@univ-lyon1.fr}.}% <-this % stops a space
\thanks{N. Noury is with the University of Lyon, INL, UCBL 
{\tt\small norbert.noury@univ-lyon1.fr}.}% <-this % stops a space
\thanks{N. Abouchi is with the University of Lyon, INL, CPE Lyon 
{\tt\small abouchi@cpe.fr}.}% <-this % stops a space
\thanks{F. Jumel is with the University of Lyon, CITI, CPE Lyon 
{\tt\small fabrice.jumel@cpe.fr}.}% <-this % stops a space
\thanks{B. Massot is with the University of Lyon, INL, INSA Lyon 
{\tt\small bertrand.massot@insa-lyon.fr}.}% <-this % stops a space
\thanks{J. Saraydaryan is with the University of Lyon, CITI, CPE Lyon
{\tt\small jacques.saraydaryan@cpe.fr}.}}% <-this % stops a space

\begin{document}

\maketitle
\thispagestyle{empty}
\pagestyle{empty}

% more at http://latexcolor.com/

\definecolor{electricblue}{rgb}{0.49, 0.98, 1.0}

% !TeX spellcheck = en_US

\begin{abstract}
An increasing number of systems use indoor positioning for many scenarios such as asset tracking, health care, games, manufacturing, logistics, shopping, and security. 
Many technologies are available and the use of depth cameras is becoming more and more attractive as this kind of device becomes affordable and easy to handle.
This paper contributes to the effort of creating an indoor positioning system based on low cost depth cameras (Kinect). 
A method is proposed to optimize the calibration of the depth cameras, to describe the multi-camera data fusion and to specify a global positioning projection to maintain the compatibility with outdoor positioning systems.

The monitoring of the people trajectories at home is intended for
the early detection of a shift in daily activities which highlights disabilities and loss of autonomy.
This system is meant to improve homecare health management at home for a better end of life at a sustainable cost for the community.
\end{abstract}

\section{Introduction}
\label{part:introduction}

With aging, a subject is likely to suffer multiple chronic diseases and a reduction in physical activity. The combination of these two effects induces a reduction in daily activities, progressive loss in autonomy, and eventually inability for an independent living if appropriate adaptations are not foreseen. A strong relationship between health status, activity, and autonomy was showed \cite{6943949}. Furthermore, the autonomy of the elderly subject relies on his ability to perform the basic actions involved in his daily living. 
Thus there is a real interest in tracking his daily activity together with the main physiological parameters (i.e. cardiac frequency, blood pressure, weight, etc.) directly at home to provide an efficient health monitoring \cite{noury2011smart}.

Since most human activities at home are attached to rooms and to interactions with equipments (i.e. to use the shower in the bathroom, to sleep on the bed in the bedroom, etc.), an activity monitoring system must be first able to describe the position of the subjects relatively to rooms and tools. After a state of the art of existing indoor positioning system, the designed system based on low cost depth vision will be described, validated over a series of tests and discussed.

\subsection{Problem constraints}

As localization is mainly descriptive in terms of space (in which room) and interaction (with which equipment or person), we shall consider that a minimum resolution of half a meter is needed since it matches an human reach.
Furthermore, since several people can be at the same time in the apartment, the localization system must be able to separate several users (possibly 5) in order to differentiate their individual activities whether these people are collaborating or acting independently. 
Multi-user tracking based on ubiquitous sensors positioned everywhere in the house has already been performed \cite{star-tracking-2005}. This paper proposes a system using less sensors while being more precise on the position an enabling the detection of social interaction.
In addition, the system must remain low intrusive and invasive in order to reduce the impact on the daily activities of the subjects.
Finally, the system cost must remain limited.
Although this system is a proof of concept, not yet fulfilling this last limitation, once produced at large scale, it shall be less expensive than renting a room in a nursing home.

\begin{table}[t]
\vspace{5pt}
 \begin{center}
\caption{Design constraints summary}
\label{tab:constraints}
\begin{tabular}[c]{cc}
\hline
%\hline
\multirow{2}{*}{\textbf{technical constraints}} 
& spacial resolution of half a meter \\
& localize several people simultaneously \\
\hline
%\hline
\multirow{3}{*}{\textbf{ergonomic constraints}}
%\hline
& low invasive  \\
& low intrusive \\
& deployment in existing flat \\
\hline
%\hline
\multirow{1}{*}{\textbf{economic constraints}}
%\hline
& lower cost than a nursing home \\
\hline
%\hline
 \end{tabular}
 \end{center}
 \vspace{-5pt}
\end{table}

\subsection{Comparison of existing indoor positioning systems}

Once those specifications were established (table \ref{tab:constraints}), several location tracking technologies had to be compared to select the best choice. 
A survey of active and passive indoor localization systems \cite{deak2012survey} classifies those technologies in two main branches: systems including an embedded wearable device and systems without, which has a great impact on the acceptability.

Some of the embedded solutions such as Ubisense \cite{steggles2005ubisense}, Ekahau RTLS \cite{kolodziej2006local} and Active Bats \cite{want1992active} 
from this survey are compliant with most of our specifications; but these are invasive and hence will be avoided if possible, even if it would be an advantage for multiple users separation and identification. Moreover, the embedded device is almost always active and thus powered on. This characteristic implies the need for a battery and a limited lifetime. A last disadvantage of these embedded systems is the inability to locate visitors who do not have the embedded device, which is quite limiting in our case.

\begin{table*}[t]
\vspace{5pt}
 \begin{center}
  \caption{Comparison of indoor positioning systems adapted from \cite{deak2012survey}}
 \label{tab:compare-non-embedded}
    \begin{tabular}[c]{|c|c|c|c|c|c|c|}
      \hline
Positioning technologies & Resolution & Cost & Multi-person localization & Invasive & Intrusive & Deployment in existing home \\ \hline
%Passive RSSI (IEEE 802.15.4) \cite{wilson2011see} & 2 m & Low & No\\
Actimetric floor \cite{valtonen2009tiletrack} & 40 cm & Medium & Yes & No & No & No \\
Passive infrared \cite{5653529} & 50 cm & Low & Yes & No & No & Yes \\
Stereo cameras \cite{krumm2000multi} & 10 cm & Medium & Yes & No & Low & Yes \\
Depth cameras \cite{kandil2014application} & 10 cm & Low & Yes & No & Low & Yes \\
Kinect \cite{6468750} & 14 cm & Low & Yes & No & Low & Yes \\
Ubisense \cite{steggles2005ubisense}  & 30 cm & Medium & Yes & Yes & Low & Yes\\
Ekahau RTLS \cite{kolodziej2006local} &  2 m & Medium & No & Yes & Low & Yes\\
Active Bats \cite{want1992active} & 9 cm & Medium & Yes & Yes & Low & Yes\\
      \hline
    \end{tabular}
  \end{center}
  \vspace{-5pt}
 \end{table*}

On the other side, among the systems without any embedded part, 
a special floor composed of many tiles \cite{valtonen2009tiletrack} can use physical contact to track people. The capacitance between floor tiles changes when someone walks on a tile. This system resolution is good enough but it is very expensive and complex to install in a pre-existing apartment.
A second solution, widely used in smart homes, including to study circadian cycles \cite{893857} is based on passive infrared sensors (PIR) which are very affordable. These sensors perform well when tracking one user but not when several users are close to each other.
Computer vision \cite{krumm2000multi} based on stereo color cameras can get a depth image of the room and localize people. Unfortunately, the depth image resolution does not allow to differentiate two people close to each other, especially in low light.
A similar approach was used to track construction workers \cite{kandil2014application} using a more advanced device: a Kinect \cite{website:kinect-ifixit}. 
%The data fusion with two Kinects have also been performed \cite{6468750},
The data fusion with two Kinects has been performed earlier \cite{6468750},
but the cameras have to be parallel and the system can only localize two people at the same time.
The Kinect is mainly composed of a color camera, a depth camera, and several microphones. 
Compared to the stereo cameras, the Kinect depth camera gives the resolution needed to localize and separate up to six people close to each other.
A positioning system based on several Kinect depth cameras is non invasive, low intrusive, and can be installed in a pre-existing apartment with the provision of accessing electrical power plugs.
Additionally, the choice of the Kinect is a low cost approach; i.e. 
an existing and widely spread technology is reused in a way which was not planned initially (it was designed as a motion sensor for games) providing the support of a large community and limiting the device price. 
Existing libraries can also be used to remove the cost of developing associated software.
For all these reasons, this passive location tracking system based on Kinect depth cameras was chosen for our indoor location tacking system. 
The positioning systems comparison is summarized in table \ref{tab:compare-non-embedded}.

\section{Materials and Methods}
\label{part:material-methods}

\subsection{The Sensor}
\label{part:sensor}

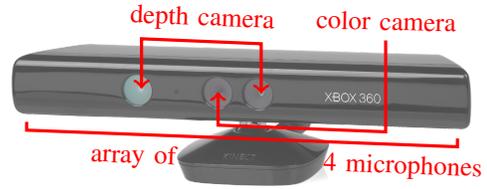
\begin{figure}[t]
\vspace{5pt}
\begin{center}
\begin{tikzpicture}
\node[anchor=south west,inner sep=0] (image) at (0,0) {
  \includegraphics[width=2.5in]{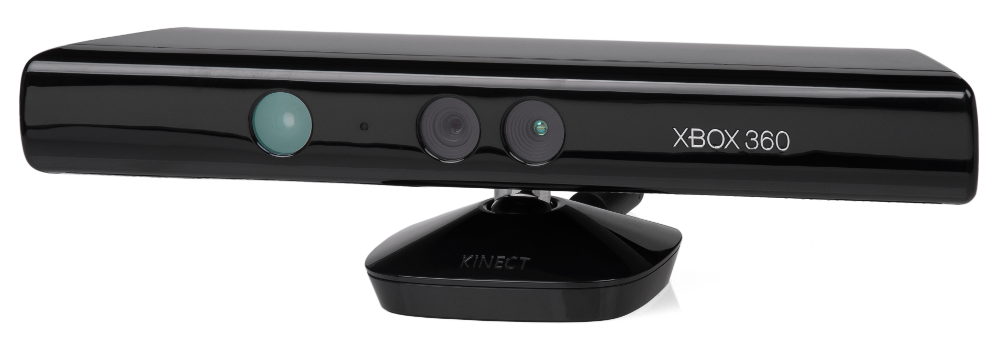}};
\begin{scope}[x={(image.south east)},y={(image.north west)}]
  \fill[opacity=0.3, white] (0,0) rectangle (1,1);
  \draw[red, very thick] (0.05,0.45) -- ++(0,-0.05) -- ++(0.9,-0.1) node[near start, below, sloped] {array of} node[very near end, below, sloped] {4 microphones} -- ++(0,0.05);
  \draw[red, very thick, <->] (0.29,0.65) -- ++(0,0.3) -- ++(0.25,-0.03) node[midway, above, sloped] {depth camera} -- ++(0,-0.3);
  \draw[red, very thick, <-] (0.45,0.63) -- ++(0,-0.2) -- ++(0.35, -0.04) -- ++(0,0.55) node[above, rotate=-1.5] {color camera};
\end{scope}
\end{tikzpicture}
\caption{Kinect sensors}
\label{fig:kinect}
\end{center} 
\vspace{-5pt}
\end{figure}

The Kinect is a motion sensor, designed by Microsoft Inc. and released in November 2010 to be used as a remote control for the Xbox 360 video game console. It is composed of several sensors (Fig. \ref{fig:kinect}): a color camera (Color CMOS, VNA38209015), a depth camera composed of an IR projector (OG12/0956/\-D306/JG05A) combined to an IR camera (IR CMOS, Microsoft/\-X853750001/\-VCA379C7130), and an array of four microphones. The infrared camera and the color camera have a definition of $640\times 480$ pixels. The angle of view is $60^{\circ}$ horizontal and $45^{\circ}$ vertical.
The device is designed to be used in a range of 0.5 m to 5 m from it. 

\subsection{Positioning in a unified landmark}

Each Kinect is directly connected to a computer running the OpenNI software \cite{website:simple-openni}.
OpenNI localizes the center of mass of the people in the camera's field of view. These 3D coordinates are in the Kinect's own landmark. 
We use several Kinects to cover  the whole flat.
Hence, the trajectories of the same person seen by several depth cameras must be compared and then merged. 
For this purpose, a projection on a unified landmark is needed. A reference landmark usable in the whole France is selected : the Lambert zone II (EPSG:27572) \cite{website:epsg-io-lambert-2}. This projection grants the ability to compare the trajectories while maintaining the compatibility with outdoor positioning systems like the GPS.

OpenNI extracts the position of a body's center of mass from the images of the depth camera. Thus, if a person is partially hidden (i.e. behind a desk), his center of mass is wrongly located along the vertical axis.
For this reason, the 3D positioning has been downgraded to a 2D positioning and the vertical axis has been neglected to remove measurement errors while staying compliant with the previously defined specifications.

Considering one of the depth cameras landmarks, a two axes coordinates must be converted in another two axes coordinates. A simple and efficient linear interpolation is used here.
The position, orientation, and tilt of the Kinect are not needed, which is a great advantage as these measures are not easily accessible. However, those Kinect positioning data influence the linear interpolation coefficients and can be retrieved from these coefficients.

Considering the person 2D position from the depth camera point of view is ($x_{cam}$, $z_{cam}$), and the position in the unified landmark is ($x_{unified}$, $y_{unified}$), the equation are:
$$x_{unified} = \alpha_1 x_{cam} + \alpha_2 z_{cam} + \alpha_3 $$
$$y_{unified} = \beta_1 x_{cam} + \beta_2 z_{cam} + \beta_3 $$

Hence, six coefficients must be computed, which means that six independent equations are needed.
Knowing one position coordinates in both landmark gives two equations.
Thus, knowing three independent positions on both landmarks gives the six needed equations and coefficients.
Those positions are named calibration points.

They are initially positioned in the reference landmark. Then, the equivalent location in the Kinect landmark is measured using OpenNI.

\subsection{Merging trajectories}

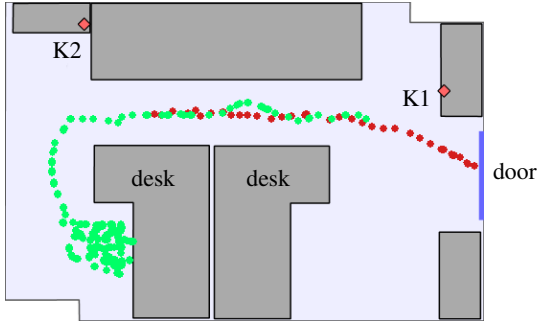
\begin{figure}[t]
\vspace{5pt}
\centering
\begin{tikzpicture}
\node[anchor=south west,inner sep=0] (image) at (0,0) {
  \includegraphics[angle=90,width=2.5in]{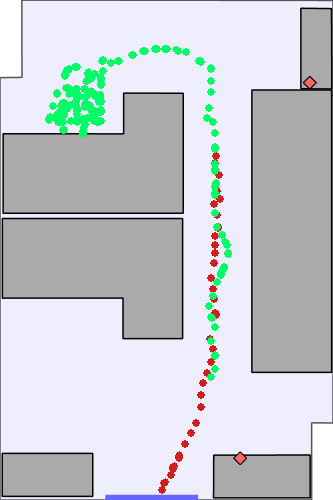}};
\begin{scope}[x={(image.south east)},y={(image.north west)}]
  %\draw[red,ultra thick,rounded corners] (0.62,0.65) rectangle (0.78,0.75);
  \node at (0.13,0.85) {\small K2};
  \node at (0.86,0.7) {\small K1};
  \node at (0.31,0.45) {\small desk};
  \node at (0.55,0.45) {\small desk};  
  \node[anchor=west] at (1,0.47) {\small door};
\end{scope}
\end{tikzpicture}
\caption{Projected trajectories from two kinects, K1 (bright green trace) and K2 (dark red trace) in an office}
\label{fig:raw-trajectory}
\vspace{-5pt}
\end{figure}

Several depth cameras are needed to cover the area of an apartment.
Hence, the system must pair the trajectories seen by two different Kinects but referring to the same person.
An example of a person successive positions, projected in the unified landmark previously defined, is shown in Fig. \ref{fig:raw-trajectory}.
In the pictured room, the two Kinects are represented as diamonds K1 and K2.
K1 is pointing to the left of the map and K2 to the right of the map.
Someone entered the room and walked through the room from the door on the right to the other side, going around the desks in the center. 
One can see in bright green and dark red the trajectories of this person detected by the two Kinects, K1 and K2 respectively.
Since K1 is pointing to the left, it cannot see the person entering the room, whereas K2 can.
Thanks to the projections on a unified landmark, 
a clear link between the two trajectories can be established (Fig. \ref{fig:raw-trajectory}).

Next, the two previous trajectories must be automatically paired in order 
to merge them and recover the full trajectory of the studied subject.
To pair trajectories, the room's area common to both Kinects field of view was used. 
In this area, the positions measured by the two Kinects were compared. 
If the correlation between the two trajectories is high enough, these may be merged in one.
As will be detailed in section \ref{part:testing}, a threshold on the correlation factor is not sufficient to decide when to merge trajectories, but it removes the impossible matches. Then, a comparison between the correlation factors provides the right peering operations. 

The correlation $C_i$ between two points of the trajectory is the product of a quality factor $Q_i \in [0,1]$ and a distance factor $D_i\in ]-\infty,1]$.
The measurement precision being variable along the Kinect's field of 
view, the quality factor includes the quality of the two measures which corresponds to the minimum distance to the calibration points or their barycenter. 
The quality of a measure is calculated as follows:
$$ Q_{i~measure} = 1-min(d_{to~calibration~points}, d_{max})/d_{max}$$
where the distances are in meters. $d_{max}$ can be adapted to the environment. It was set to 2 m in this room from our experimental data.

Also, the two measures are not sampled at the exact same time. Hence the time delay between the two measures, is included in the quality factor calculation:
$$Q_{i~time} = max(0,1-2\Delta t^2)$$
Where $\Delta t$ is in seconds.

Then the quality factor for an association of two samples becomes:
$$ Q_{i} = Q_{i~measure~1} \times Q_{i~measure~2} \times Q_{i~time} $$
As explained above, $Q_i \in [0,1]$ where $1$ is the best quality.
The distance factor $D_i$ is calculated as follow: 
$$D_i=(1-d_{between~points}^2)$$ 
where the distance is in meters. 
Consequently, $D_i$ is close to $1$ when the two measured positions are close and becomes negative when they are further away from each other.
Eventually, the correlation $C$ between the two trajectories is the sum of the $C_i$.
$$C=\sum_i C_i = \sum_i D_i\times Q_i$$ 
Close positions with good quality factor will result in an increase of the correlation sum. On the contrary, few samples showing that the two positions are not corresponding at all will decrease the correlation sum to a negative value.
The threshold of the correlation sum to consider merging trajectories was set empirically from our experimental data. 
If the correlation between two trajectories does not reach the threshold, those cannot be merged. Otherwise, the correlation factors are compared and the highest one is chosen for the trajectory peering.

\section{Experimentation and Validation}
\label{part:testing}

\subsection{Selecting the calibration points}

\begin{figure}[t]
\vspace{5pt}
\centering
\begin{tikzpicture}
\node[anchor=south west,inner sep=0] (image) at (0,0) {
  \includegraphics[width=2.5in]{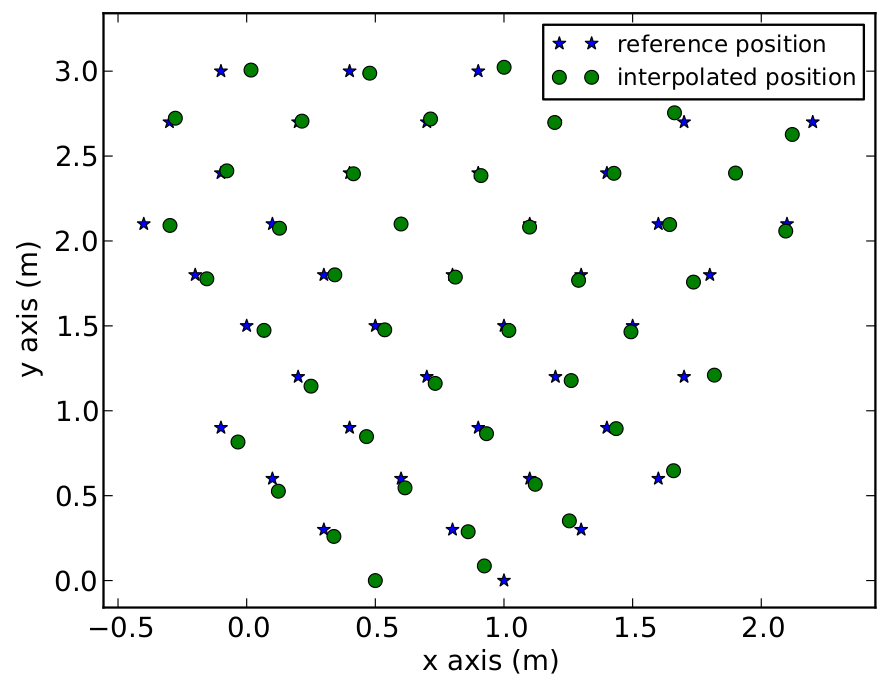}};
\begin{scope}[x={(image.south east)},y={(image.north west)}]
  %\draw[red,ultra thick,rounded corners] (0.62,0.65) rectangle (0.78,0.75);
  %\node at (0.13,0.85) {\small K2};
\end{scope}
\end{tikzpicture}
\caption{Reference grid}
\label{fig:grid}
\vspace{-5pt}
\end{figure}

The selection of calibration points is of major importance to create the projection of each Kinect landmark in the unified landmark. 
Hence, the error introduced by this projection partly rely on the choice of this calibration points. 
In order to find out the best way to select the three calibration points, a comparison was made. 
This experience uses a grid on the floor and one depth camera. The grid is composed of forty-seven points regularly positioned, between one and five meters to the camera. 
A person stands successively on each point of the grid and his position relatively to the depth camera is measured using OpenNI. 
As in the calibration process, for each point of the grid, the coordinates are known in both landmarks. 
Hence, any set of three points can be used to do the projection from the Kinect landmark to reference one. 
For every set of three points, this projection is done. 
One of the projection results is shown in Fig. \ref{fig:grid} where the reference positions are stars and the interpolated positions from the depth camera data are circles. 

The mean error (the distance between the reference point and the interpolated one) for all other points is measured. 
Depending on the three points selected to make the linear interpolation, the mean error can be lower than half a meter, but it can also be above  three meters. Hence, an indicator is needed to select the best interpolation points, without having to use this grid of points every time.

\begin{figure}[t]
\vspace{5pt}
\centering
\includegraphics[width=2.5in]{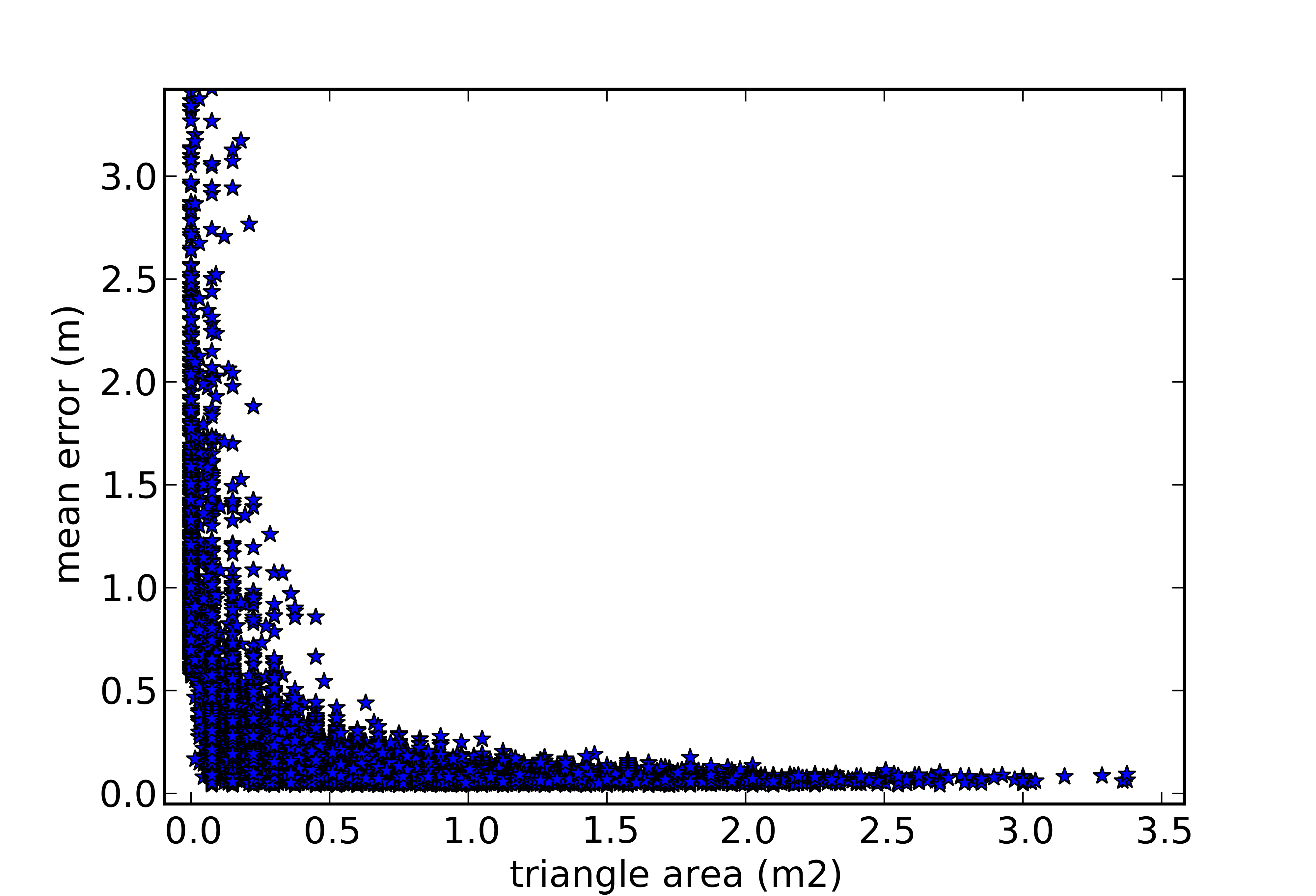}
\caption{Influence of the area between calibration points}
\label{fig:influence-area}
\vspace{-5pt}
\end{figure}

For every set of three calibration points, the measured mean error was compared to the area of the triangle created by these three calibration points. 
The result is shown in Fig. \ref{fig:influence-area}. 
The influence of the area is clear. 
The wider the area is, the lower the upper bound will be.
This demonstrates, we can reduce the error introduced by the interpolation by maximizing the area formed by the three selected calibration points. 

\begin{figure}[t]
\vspace{5pt}
\centering
\includegraphics[width=2.5in]{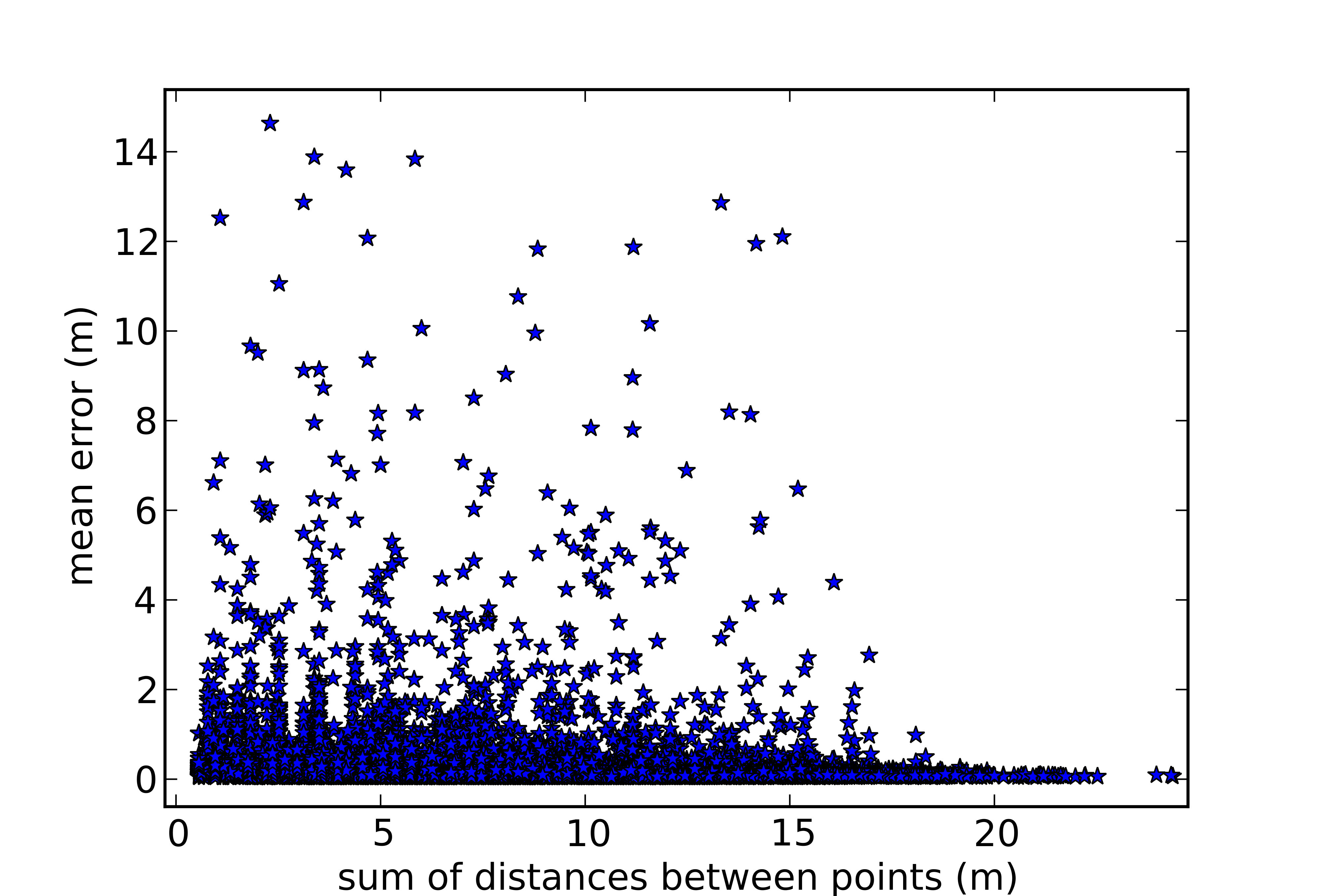}
\caption{Influence of the distance between points}
\label{fig:influence-distance}
\vspace{-5pt}
\end{figure}

The same comparison was made between the mean error and the perimeter of the triangle. The result is shown in Fig. \ref{fig:influence-distance}.  
The link between the perimeter and the error is not evident.
This exhibits the relevance of the area indicator to select the best calibration points in order to lower the introduced error. This link is logical since aligned points would introduce an important error in the perpendicular direction and since the further the points are, the lower the measurement error bias should be.

In the office used for the test, the area of the triangle formed by the three calibration points of the depth camera K1 is $2.5~m^2$ and for K2, $1.5~m^2$. Hence, the mean error should be lower than $0.2~m$ is both cases. This result also validates the projection method as it respects our initial specifications of a spacial resolution below the half-meter.

\subsection{Ability to separate two people}

\begin{figure}[t]
\vspace{5pt}
\centering
\includegraphics[width=2.5in]{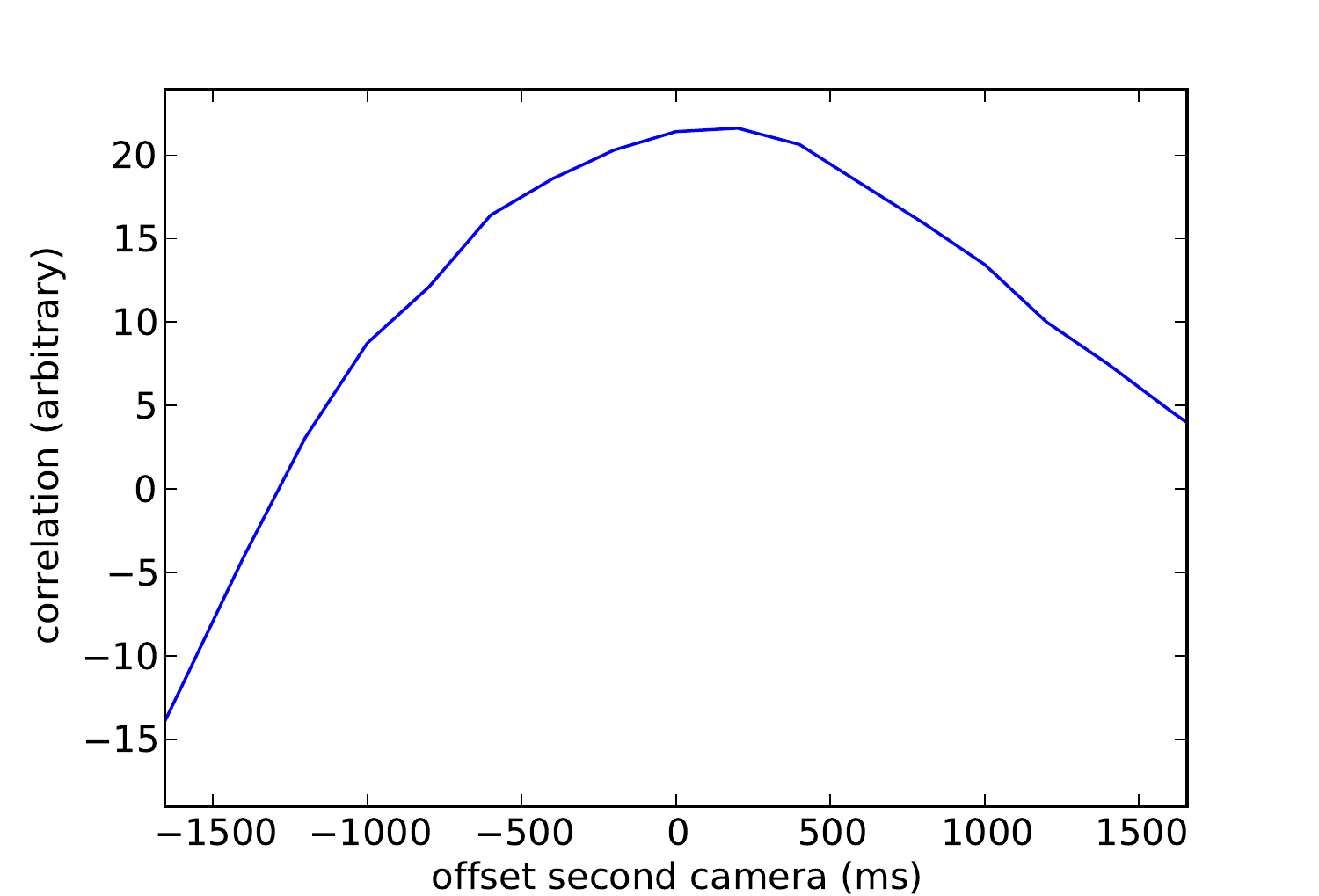}
\caption{Influence of a time offset on a camera}
\label{fig:influence-time}
\vspace{-5pt}
\end{figure}

As proposed in introduction, this system must be able to separate the trajectories of two people following each other.
To validate this capacity, a time offset was introduced on a sampled trajectory in order to simulate two people following exactly the same path which would be the worst case. The results are shown in Fig. \ref{fig:influence-time}.

When the offset is one second, the correlation drops down from $21.4$ to $13.5$. This result shows that, comparing the correlations, the trajectory merging method is able to separate two people following each other closely (one second offset is equivalent to about one meter at normal walking speed).

\section{Discussion}

The proposed location tracking system is based on existing depth camera and open source software to lower the cost, ensure sustainability, and benefit from the community collaboration.
The low cost approach, including the deviation of the initial use of the Kinect, gave the ability to have an operative system much quicker than with a full development from scratch with well reduced expenses.
No equipment has to be worn by the tracked people. This way, the system is even able to locate unknown visitors. The drawback of using several Kinect depth cameras is the need of one computer per camera and thus, of a power line everywhere a Kinect is installed.

The tests performed on one experimental trajectory seen by two depth cameras are encouraging. 
Those tests guided the choice of the empirical threshold used in the trajectory merging process. 
Yet, a larger set of scenarios is needed in order to fully validate the capacity of the system when facing several kinds of activity.
The area common to two cameras will also vary as more rooms are equipped and the threshold will have to be adapted to these parameters.

The images taken by the cameras are always processed on the computer linked to the Kinect, never stored or transfered over the network. 
This should reassure users on potential loss in privacy,
even if some may still consider the cameras intrusive.

Moreover, being able to track several users in an apartment impose the ability to identify the detected people. 
Otherwise, it would be impossible to associate the same trajectory to someone going in and out the apartment. 
Hence, an identification module must be added to the location tracking system. 
As a first step, the color camera available in the Kinect can be used, but it needs an operator to identify the pictures taken with the color camera. Image processing systems for face recognition exist but the constraints in terms of image quality and orientation cannot be reached with a Kinect camera. Hence, an autonomous identification module, based on the RFID technology, will be further developed and integrated in the system.

\section{Conclusion}

The indoor location tracking system presented in this paper is operational.
This low cost system is easy to deploy, very modular, and able to evolve with technologies. 
It is non invasive since no embedded part is worn by the subject and does not require batteries. 
The fusion of positioning data is efficient, being able to differentiate individuals following the same trajectory. 

By locating people relatively to rooms and equipment, the system provides an efficient way to measure, understand, and monitor people activity along the days.
The next step will be to complete this location process with the identification process of the people living in the apartment, and to recognize a set of life scenarios, leading to a personalized daily activity tracking system.
This will enable a better understanding of people's life cycles like circadian ones as experimented earlier \cite{6943949}, of people's needs, and to sustain a longer autonomy at home for the elderly.

\addtolength{\textheight}{-10cm}   % This command serves to balance the column lengths

\section*{ACKNOWLEDGMENT}
The authors would like to thank the Institute of Nanotechnologies of Lyon (INL) for the special grant attributed to Lo\"ic Sevrin for his PhD thesis.

%%%%%%%%%%%%%%%%%%%%%%%%%%%%%%%%%%%%%%%%%%%%%%%%%%%%%%%%%%%%%%%%%%%%%%%%%%%%%%%%

\bibliographystyle{IEEEtran}
\bibliography{ref2}

\end{document}